\documentclass[twocolumn,nofootinbib,amsmath,prd,aps,superscriptaddress,tightenlines,preprintnumbers,nobalancelastpage]{revtex4-1}

\usepackage{graphicx}
\usepackage{braket}
\usepackage[usenames]{color}
\usepackage{latexsym}
\usepackage{amsmath}
\usepackage{mathrsfs}
\usepackage{hyperref}
\usepackage{enumerate}
\usepackage{adjustbox}
\usepackage{amssymb}
\usepackage{slashed}

\def\bea{\begin{eqnarray}}
\def\eea{\end{eqnarray}}

\begin{document}
\newcount\hour \newcount\minute
\hour=\time \divide \hour by 60
\minute=\time
\count99=\hour \multiply \count99 by -60 \advance \minute by \count99
\newcommand{\mydate}{\ \today \ - \number\hour :00}

\preprint{IPPP/18/19}
\title{On the behaviour of composite resonances\\[0.5cm] breaking lepton flavour universality}

\author{Mikael Chala and Michael Spannowsky\\\vspace{0.4cm}
\it {Institute of Particle Physics Phenomenology, Physics Department, Durham University, Durham DH1 3LE, UK}
}

\begin{abstract}
Within the context of composite Higgs models, recent hints on lepton-flavour non-universality in $B$ decays can be explained by a vector resonance $V$ with sizeable couplings to the Standard Model leptons ($\ell$). We argue that, in such a case, spin-$1/2$ leptonic resonances ($L$) are most probably light enough to open the decay mode $V \rightarrow L\ell$. This implies, in combination with the fact that couplings between composite resonances are much larger than those between composite and elementary fields, that this new decay can be important. In this paper, we explore under which conditions it dominates over other decay modes. Its discovery, however, requires a dedicated search strategy. Employing jet substructure techniques, we analyse the final state with largest branching ratio, namely $\mu^+\mu^- Z/h, Z/h \rightarrow$ jets. We show that \textit{(i)} parameter space regions that were believed excluded by di-muon searches are still allowed, \textit{(ii)} these regions can already be tested with the dedicated search we propose and \textit{(iii)} $V$ masses as large as $\sim 3.5$ TeV can be probed at the LHC during the high-luminosity phase.
\end{abstract}

\maketitle
\newpage

\section{Introduction}
Experimental data collected during the last few years by LHCb~\cite{Aaij:2013qta,Aaij:2014ora,Aaij:2014pli,Aaij:2015esa,Aaij:2015oid,Aaij:2017vbb}, Belle~\cite{Wehle:2016yoi} and the LHC~\cite{Khachatryan:2015isa,Sirunyan:2017dhj} suggest departures from Lepton Flavor Universality (LFU) in $B$ meson decays with respect to Standard Model (SM) predictions. In particular, the measured values of the very clean observables
\begin{equation}
R_{K^{(*)}} \equiv \frac{\mathcal{B}(B^{+(0)}\rightarrow K^{+(*)}\mu^+\mu^-)}{\mathcal{B}(B^{+(0)}\rightarrow K^{+(*)}e^+e^-)}~
\end{equation}
depart from the SM prediction~\cite{Hiller:2003js,Bordone:2016gaq} by more than $2\,\sigma$, while a naive combination results in a discrepancy of about $4\,\sigma$~\cite{Capdevila:2017bsm}. 

Among other possibilities, it has been proposed that the origin of LFU violation relies on a composite spin-1 resonance $V$ with sizeable couplings to the SM leptons~\cite{Niehoff:2015bfa,Niehoff:2015iaa,Carmona:2015ena,Megias:2016bde,GarciaGarcia:2016nvr,Megias:2017ove,Sannino:2017utc,Carmona:2017fsn}; see also Ref.~\cite{King:2017anf}. This kind of particle arises naturally in composite Higgs models (CHMs)~\cite{Kaplan:1983fs,Kaplan:1983sm}, which are further motivated by the gauge-hierarchy problem. For concreteness, we will focus on the muon case, but most of the discussion and results can be extended straightforwardly to  electrons~\cite{Carmona:2015ena,Carmona:2017fsn}. 

As we will argue, this solution implies that vector-like partners $(L)$ of the SM leptons ($\ell$), can very well be lighter than $V$. In that case, the channel $V\rightarrow L\ell$ opens and it can actually dominate the $V$ decay width. It is certainly surprising that no single study has taken this effect into account. This has two major implications. \textit{(i)} Contrary to the standard case studied so far, this setup survives all constraints from LHC data, including the strongest ones from di-muon searches. \textit{(ii)} In already collected events containing two muons and a fat jet resulting from a $Z$ or Higgs boson, new dedicated searches can reveal a clear peak in the invariant mass distribution of $m_{\mu_1 j_1} = \sqrt{(p_{\mu_1} + p_{j_1})^2} $, with $\mu_1$ and $j_1$ being the highest-$p_T$ muon and jet, respectively.

%
\section{Model}~\label{ref:model}
The phenomenological Lagrangian describing the interactions between the spin-1 singlet $V$ and the SM leptons is given by
\begin{equation}
\Delta\mathcal{L} = \frac{1}{2}m_V^2 V_\mu V^\mu + J_\mu V^\mu + \cdots
\end{equation}
where $m_V$ is the mass of $V$ and the ellipsis encode the kinetic term as well as other interactions not relevant for the subsequent discussion. We further define
\begin{equation}\label{eq:par}
J_\mu = g_{V\ell\ell} \lambda^\ell_{ij}\overline{\ell_L^i} \gamma_\mu\ell_L^j + g_{Vqq} \lambda^q_{ij}\overline{q_L^i} \gamma_\mu q_{L}^j~,
\end{equation}
with $\ell$ and $q$ SM leptons and quarks, respectively.
Let us consider for simplicity $\lambda_{ij}^\ell \sim \delta^2_i \delta^2_j$ and $\lambda_{ij}^q \sim \delta^3_i \delta^3_j$, so that mainly the second-generation leptons and the third-generation quarks couple to the vector resonance. Likewise, the LFU violating term arises in the physical basis after performing the CKM rotation in the down sector, and can therefore be estimated as $\sim g_{Vqq}V_{ts}^{\mathrm{CKM}}V_{tb}^{\mathrm{CKM}}$. Reproducing the LFU anomalies requires~\cite{Crivellin:2015era,DiLuzio:2017fdq}%
\begin{equation}\label{eq:fit}
 g_{Vqq} \sim 0.05 \frac{m_V^2}{\text{TeV}^2}
\end{equation}
for $g_{V\ell\ell}\sim 1$. Much larger values of $g_{V\ell\ell}$ are disfavoured by limits on neutrino trident production~\cite{Alok:2017jgr}. On the other hand, smaller values are disfavoured by measurements of $\Delta M_s$ for values of $m_V$ in the natural region of CHMs, namely $m_V \sim $ few TeV. We thus stick to this value henceforth, which is allowed even by the latest measurement of $\Delta M_s$~\cite{DiLuzio:2017fdq}.
The key point is that, if $V$ is a resonance in a CHM, its couplings to the SM fermions originate from partial compositeness~\cite{Kaplan:1991dc}. The more fundamental Lagrangian reads 
\begin{align}~\label{eq:simp}\nonumber
\Delta \mathcal{L} &= \frac{1}{2} \frac{M_{V}^2}{g_c^2} (g_c V^\mu - g_e B^\mu)^2 \\\nonumber
&+ \overline{L}(i\slashed{D} - g_c\slashed{V} - M_L) L + \overline{Q}(i\slashed{D} - g_c\slashed{V} - M_Q) Q  \\
&+   \bigg[\Delta_{L}\overline{\ell_L} L + \Delta_{Q}\overline{q} Q + \text{h.c.}\bigg] +  \cdots
\end{align}
where $\slashed{V} = \gamma_\mu V^\mu$, $g_e$ ($g_c$) is a weak (strong) coupling, $\slashed{D}$ is the SM covariant derivative, $M_{L,Q}, \Delta_{L,Q}$ are dimensionful constants, $Q = (T~ B)^t$ and $L = (E~ N)^t$ are composite $SU(2)_L$ fermion doublets and $V$ and $B$ are composite and elementary vectors, respectively. The physical vectors are admixtures of the latter with mixing angle $\theta$. Likewise for the fermions. With a slight abuse of notation, we denote the physical fields with the same letters.

\begin{figure}[t]
 \includegraphics[width=\columnwidth]{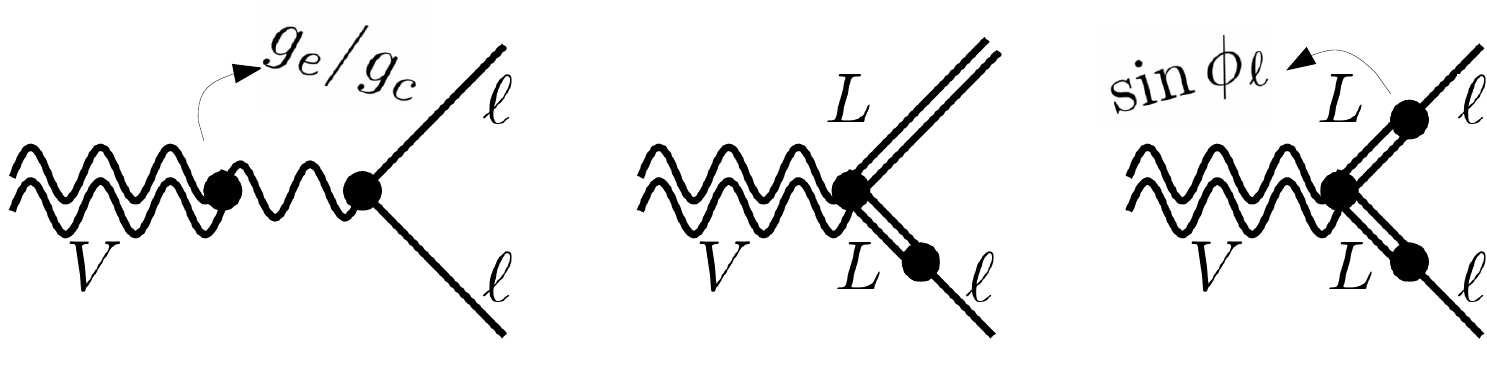}
 \caption{\it Different ways in which the composite vector $V$ interacts with the SM fermions. The left diagram is suppressed by the smallness of $g_e/g_c$. The right one is suppressed by one more power of $\sin{\phi_\ell}$ with respect to the center one.}\label{fig:diagram}
\end{figure}

After rotating the heavy and the light degrees of freedom, the following relations hold:
\begin{align}\label{eq:relations}\nonumber
\tan{\theta} &= \frac{g_e}{g_c}~, \, g' = g_e \cos{\theta} = g_c\sin{\theta}~, \, m_V = M_V\cos{\theta}~,\\
m_{L, Q} &= \frac{M_{L,Q}}{\cos{\phi_{\ell, q}}}~, \quad \tan{\phi_{\ell, q}} = \frac{\Delta_{L, Q}}{M_{L, Q}}~, 
\end{align}
with $m_{L,Q}$ the physical masses before Electroweak Symmetry Breaking (EWSB) of the vector-like fermions and $g'$ the $U(1)_Y$ gauge coupling.
In the expected limit $g_c \gg g_e$, we find
\begin{equation}\label{eq:gVff}
g_{V\ell\ell} \sim g' \sin^2{\phi_\ell} \cot{\theta}, \quad g_{VL\ell} \sim g' \frac{\sin{\phi_\ell}\cos{\phi_\ell}}{\sin{\theta}\cos{\theta}},
\end{equation}
where $g_{VL\ell}$ parametrises the strength of the $VL\ell$ interaction; see a pictorial representation in Fig.~\ref{fig:diagram}. Similar expressions hold in the quark sector. In this limit, $\cot{\theta}$ is large, the mixing between $V$ and the SM gauge bosons is small and $V$ production in the $s$-channel at proton colliders is dominated by bottom quarks.

Following Eq.~\ref{eq:fit}, we obtain
\begin{equation}
 \sin^2{\phi_{\ell}} \sim \frac{1}{g' \cot{\theta}}, \quad \sin^2{\phi_q}\sim \frac{0.05}{g' \cot{\theta}} \frac{m_V^2}{\text{TeV}^2}~.
\end{equation}
This implies that the degree of compositeness of the second-generation leptons is large, even larger than that of the left-handed quarks. 
On top of this, the top Yukawa $y_t$ is induced by proto-Yukawa interactions $\sim Y(\overline{T} H Q^\prime+\text{h.c.})$, with $Y$ a dimensionless coupling, and $T$ and $Q^\prime$ composite resonances mixing with $t_R$ and $q$ respectively. The phenomenology of $V$ depends still on the dynamics of these particles. This can not be inferred from the anomaly data. Instead, we must rely on the possible structure of the CHM. Let us discuss different regimes and their implications.

$\blacksquare$ Let us assume that $Q^\prime = Q$ and that $T$ couples also to 
$V$ with $g_c$ strength. Then $y_t\sim Y \sin{\phi_q}\sin{\phi_t}$, with 
$\phi_t$ the degree of compositeness of $t_R$. We can in turn distinguish  two 
cases:

\textit{(i)} If $Y\lesssim 4$ we get $\sin{\phi_t} \sim 1$. In this regime, $t_R$ is maximally composite, and $V$ decays predominantly to $t\overline{t}$. 

\textit{(ii)} If $Y\gtrsim 4$ then $\sin{\phi_t}$ can be significantly smaller than $\sim 1$. In that case, the $V$ decay width to SM particles is small. The dominant modes are $V\rightarrow Tt$ and $V\rightarrow L\ell$. The former is more relevant for smaller values of $Y$, $m_Q$ and $m_T$ and for larger values of $m_L$; see Refs.~\cite{Barcelo:2011wu,Bini:2011zb,Carmona:2012jk,Chala:2014mma} for dedicated analyses. The latter dominates otherwise. As an example, $\sin{\phi_t}\sim 0.5$ implies that $g_{VL\ell}\sim g_{VTt}\sim 2.5$. In such case, for $m_V = 2$ TeV, $m_L = 500$ GeV and $m_T = 1.5$ TeV we already get that $\Gamma(V\rightarrow L\ell)$ is around $1.25$ times larger than $\Gamma(V\rightarrow Tt)$.

Let us also notice that the aforementioned values for the masses fit well within the current experimental data on CHMs. Indeed, the Higgs in CHMs is an approximate Goldstone boson. Its mass is generated radiatively and grows with $\Delta_{L,Q}$, because these are the main sources of explicit breaking of the shift symmetry. Thus, in order not to advocate a large cancellation between different breaking sources to keep the Higgs light,  $\Delta_{L}$ must be small. Given the large value of $\sin{\phi_{\ell}}$, this implies a small $m_{L}$; see Eq.~\ref{eq:relations}. In addition, there is no experimental reason for $m_L$ not to be in the sub-TeV region. In fact, due to the small EW pair-production cross section for heavy vector-like leptons, they can be as light as few hundreds GeVs~\cite{Redi:2013pga}. Very dedicated searches will be needed to unravel larger masses even in the LHC High-Luminosity (HL) phase~\cite{delAguila:2010es}. 
The preferred values of $m_{T}$ (and $m_Q$), namely $ \lesssim 1$ TeV~\cite{Contino:2006qr,Matsedonskyi:2012ym,Redi:2012ha,Marzocca:2012zn,Pomarol:2012qf,Panico:2012uw,Panico:2015jxa}, are instead in tension with current LHC data, at least in the minimal CHM~\cite{Chala:2017xgc}. Even masses as large as $m_T\sim 1.3$ TeV have been already ruled out in several scenarios~\cite{Aaboud:2018saj}.

$\blacksquare$ If $Q^\prime\neq Q$ and it does not couple to $V$, or if composite right-handed currents (such as $\overline{T}\gamma^\mu T$) couple less to the spin-1 resonance, then $g_{Vtt}$ and $g_{VTt}$ can be arbitrarily small. In light of this observation, and given the discussion in point \textit{(ii)} and the fact that $t\overline{t}$ and $Tt$ signatures have already been explored in the literature, we focus on this regime hereafter. We will show that $V\rightarrow L\ell$ plays a dominant role in this case.

Finally, it is worth mentioning that other uncoloured composite vector resonances such as EW triplets, commonly present in CHMs too, couple directly to the Higgs and the longitudinal polarization of the gauge bosons~\cite{Panico:2015jxa}. Given that the latter are fully composite, non-singlet vectors decay mostly into them and not into pairs of heavy-light leptons~\cite{Agashe:2016ttz,Agashe:2017ann}

\section{Composite vector phenomenology}~\label{ref:pheno}
The leading-order decay widths for $V$ in the limit $m_V \gg m_\ell, m_q$ are
\begin{align}
\Gamma_{V\to \ell\ell} &= \frac{1}{24\pi} g_{V\ell\ell}^2 m_V~,\\ \nonumber
\Gamma_{V\to qq} &= \frac{1}{8\pi} g_{Vqq}^2 m_V~,\\ \nonumber
\Gamma_{V\to L\ell} &= \frac{1}{24\pi} g_{VL\ell}^2 m_V \left[1 - \frac{m_L^2}{m_V^2}\right]\left[1 - \frac{m_L^2}{2 m_V^2} - \frac{m_L^4}{2 m_V^4}\right]~.
\end{align}

We willl restrict ourselves to the regime $m_V < 2 m_L$. Otherwise, the decay into two heavy fermions opens and $V$ becomes typically too broad to be treated as a resonance~\cite{Chala:2014mma}. We will consider a Benchmark Point (BP) defined by $m_V = 2$ TeV, $m_L = 1.2$ TeV and $\cot{\theta} = 20$. Departures from this assumption will be also discussed. We note that, while $g_{Vqq}$ depends on $m_V$ and is weak, $g_{V\ell\ell}$ and $g_{VL\ell}$ are approximately fixed and given by $\sim 1$ and $\sim 2.5$, respectively. They are all below the perturbative unitarity limit $\sim\sqrt{4\pi}$.
Likewise, $\Gamma_{V}/m_V$ is never above $30$ \%. A perturbative approach to the collider phenomenology of $V$ is therefore justified.

We do not aim to focus on any particular UV realization of the simplified Lagrangian in Eq.~\ref{eq:simp}. Our aim is rather highlighting the implications of light lepton partners for the phenomenology of $V$. However, it must be noticed that composite muons give generally large corrections to the $Z\mu_L\mu_L$ coupling. These can be avoided in left-right symmetric implementations of lepton compositeness~\cite{Agashe:2009tu}. This requires however the introduction of more degrees of freedom. For example, second-generation leptons in the minimal CHM~\cite{Agashe:2004rs} might mix with composite resonances transforming in the representation $\mathbf{10}$ of $SO(5)$; see Ref.~\cite{Niehoff:2015bfa}\footnote{This reference showed also that this choice modifies the $Z\overline{\nu}\nu$ coupling, making the $Z$ invisible decay width fit better the observed deficit by LEP.}. The latter reduces to $(\mathbf{2},\mathbf{2})+(\mathbf{3}, 1)+(1,\mathbf{3})$ under the custodial symmetry group $SO(4)$. Interestingly, the extra degrees of freedom, namely $(1, \mathbf{3})$ and $(\mathbf{3}, 1)$, do not affect the spin-1 vector decays for several reasons.

\textit{(i)} In the regime we are interested in, pair-production of heavy leptons mediated by $V$ is kinematically suppressed. 

\textit{(ii)} The custodial triplets do not mix with the SM fermions before EWSB. (Note also that the product of $(\mathbf{2},\mathbf{2})$ times any of the custodial triplets can not be a singlet, and hence the corresponding current does not couple to $V$.) Therefore, the extra new fermions can only be produced in association with SM fermions with a strength further suppressed by a factor of $Y v/M$, $Y$ being the typical coupling between composite fermions.
Provided this is not extremely large, the extra states can be ignored~\cite{Chala:2013ega}. (Similar reasonings work for other representations.) In this regime the following relations hold with good accuracy:
\begin{align}
\mathcal{B}(E^\pm\to Z\mu^\pm) \simeq ~&\mathcal{B}(E^\pm\to h\mu^\pm)\simeq 0.5, \\
&\mathcal{B}(N\to W^\pm\mu^\mp)\simeq 1~.
\end{align}
Hereafter, we assume this to be the case.
\begin{figure}[t]
 \includegraphics[width=\columnwidth]{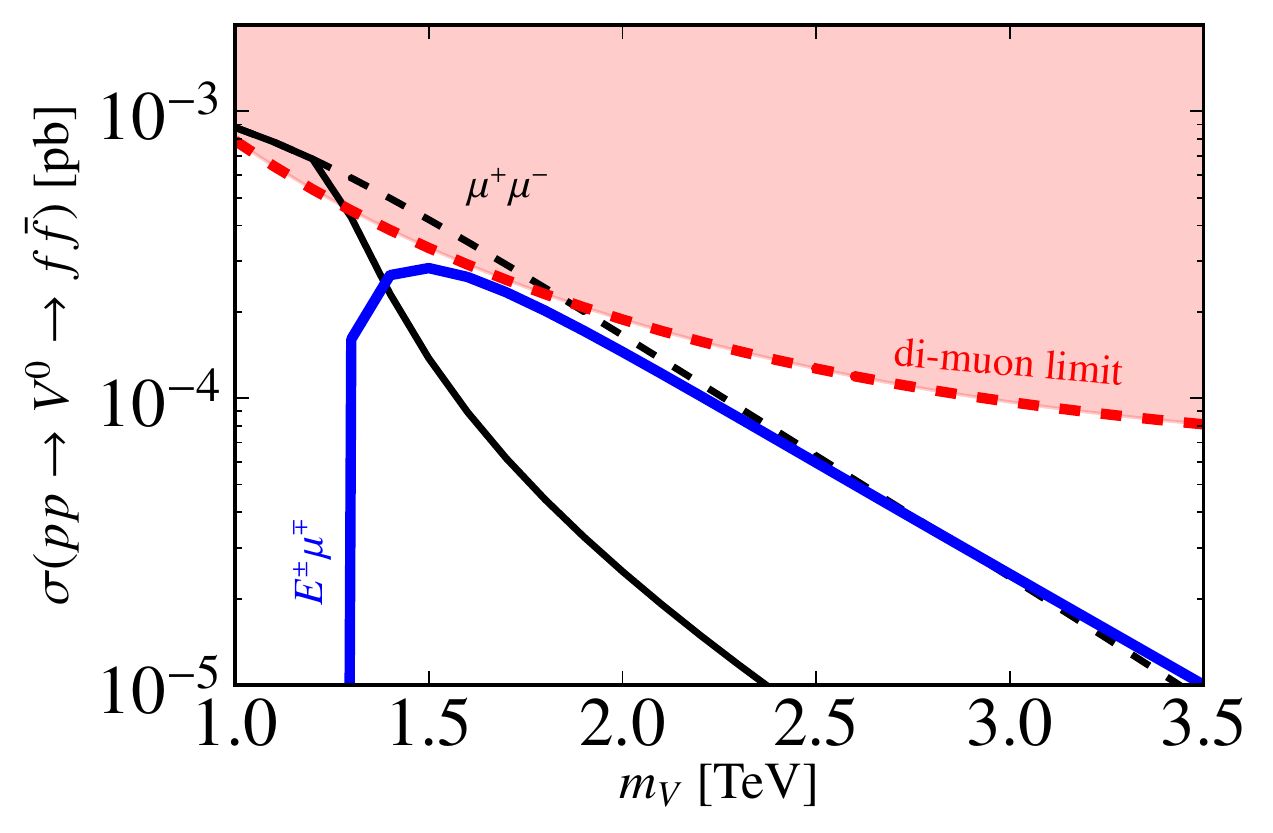}
 \caption{\it Production cross section for $E^\pm\mu^\mp$ (thick solid blue) and $\mu^+\mu^-$ (thin solid black) in the BP. The thin dashed black line represents the cross section for $\mu^+\mu^-$ for $m_L > m_V/2$. In dashed red, we see the current limits on this cross section from ATLAS~\cite{ATLAS:2017wce}.}\label{fig:limits}
\end{figure}
The production cross section for $pp\to V\to \mu^+\mu^-$ in the BP at the LHC with $\sqrt{s} = 13$ TeV is depicted by the thin black solid line in Fig.~\ref{fig:limits}. The thin black dashed line represents the would-be cross section in the absence of light $L$. The region above the thick red dashed curve is excluded according to the recent ATLAS analysis of Ref.~\cite{ATLAS:2017wce}. Clearly, in the region where the heavy-light topology is kinematically forbidden, di-muon constraints are extremely important, the limit on $m_V$ being close to $1.8$ TeV. However, in the presence of light lepton partners, and given that the coupling between composite particles is larger than that between composite and elementary fields, the cross section $pp\to V\to E^\pm\mu^\mp$ dominates (thick blue solid line). The limit on $m_V$ imposed by di-muon searches gets then reduced by more than $500$ GeV. 
Additionally, dijet searches as well as $t\overline{t}$ searches are less constraining.
Other LHC searches are sensitive to the heavy-light channel.
In particular, we considered searches for electroweakinos in multilepton events with large missing energy, such as that in Ref.~\cite{ATLAS-CONF-2017-039}. The 13 fb$^{-1}$ version of this analysis was first presented in Ref.~\cite{ATLAS-CONF-2016-096}. The latter is fully included in \texttt{CheckMATE v2}~\cite{Dercks:2016npn}. Therefore, the reach of the former can be estimated by scaling the signal over the $95\%$ CL limit by the luminosity ratio $\sqrt{36/15}$. We obtain that values of $m_V$ above $1$ TeV are not constrained.

Other studies, such as searches for evidence of the type-III seesaw mechanism~\cite{Sirunyan:2017qkz}, focus on final states with more than two leptons, which are almost absent in our scenario. On balance, we find imperative to develop a dedicated search to unravel the origin of LFU in CHMs.

\section{New searches}~\label{ref:analysis}
%
%
Among the different heavy-light topologies, we focus on the channel $pp\to V\to E^\pm\mu^\mp, E^\pm\to Z/h ~ \mu^\pm$, with $Z/h$ decaying hadronically\footnote{The leptonic $Z$ channel, although extremely clean, produces less than $10$ events before cuts for $m_V > 2$ TeV even with a luminosity $\mathcal{L} = 3000$ fb$^{-1}$.}. 
%

Signal events are generated using \texttt{MadGraph v5}~\cite{Alwall:2014hca} and \texttt{Pythia v6}~\cite{Sjostrand:2006za}, after including the relevant interactions in an \texttt{UFO} model~\cite{Degrande:2011ua} using \texttt{Feynrules v2}~\cite{Alloul:2013bka}. For the subsequent analysis, we have used home-made routines based on \texttt{Fasjet v3}~\cite{Cacciari:2011ma} and \texttt{ROOT v6}~\cite{Brun:1997pa}. Muons are defined by $p_T > 20$ GeV and $|\eta| < 2.7$. Jets are clustered according to the Cambridge-Aachen algorithm~\cite{Dokshitzer:1997in,Wobisch:1998wt} with $R = 1.2$. Muons with $p_T > 50$ GeV are removed from hadrons in the clustering process. The dominant background is given by $\mu^+ \mu^- +$ $\text{jets}$. We matched Monte Carlo background events with $1$ and $2$ jets using the \texttt{MLM} merging scheme~\cite{Hoche:2006ph} with a matching scale $Q = 30$ GeV. At the generator level, we also impose a cut on the $p_T$ of the muons, $p_T^\mu > 100$ GeV. The matched cross section we obtain at LO at $\sqrt{s} = 13$ TeV is $\sim 1.2$ pb; we generated $10$ million events.

As basic cuts we require, first, the presence of exactly two opposite charged muons and at least one jet. The leading $p_T$ jet, $j_1$, is required to have a significant mass drop~\cite{Butterworth:2008iy}, characterized by $\mu = 0.67, y_\text{cut} = 0.3$. This jet is further filtered~\cite{Butterworth:2008iy, Soper:2010xk} using a finer angular scale given by $R_\text{filt} = \text{min}\lbrace 0.3, 0.5 \times R_{12}\rbrace$, with $R_{12}$ the angular separation of the two sub-jets obtained in the mass-drop procedure. This method impacts on the background by systematically moving $m_{j_1}$ to smaller values. Likewise, the $h$ and $Z$ boson mass peaks in the signal become significantly narrower.
We also impose both muons to have a $p_T > 200$ GeV. This stringent cut is motivated by the fact that muons originate from the decay of very heavy particles, while the background is mostly populated by soft leptons. More sophisticated jet substructure methods can improve on our result further \cite{Soper:2011cr, Soper:2012pb, Adams:2015hiv, Larkoski:2017jix}.

Finally, we enforce the leading jet to have an invariant mass $80~\text{GeV}\leq m_{j_1} \leq 130$ GeV. A summary of the basic selection cuts is given in Tab.~\ref{tab:cuts}. Their efficiency in the BP as well as in the background are also displayed. Interestingly, while a large fraction of the signal is kept, the background is reduced by more than two orders of magnitude.

\begin{table}[ht]
\begin{center}
\begin{adjustbox}{width = 0.9\columnwidth}
\footnotesize
\begin{tabular}{|l|l|l|}\hline
 &  $\epsilon(BP)$ & $\epsilon(b)$ \\\hline
$2$ muons & $90$ & $99$   \\
$\geqslant 1$ jet, $j_1$ tagged and filtered & $70$ & $45$  \\
$p_T^{\mu_{1,2}} > 200$ GeV & $93$ & $16$  \\
$80$ GeV $< m_{j_1} < 130$ GeV & $58$ & $7.0$ \\\hline
Total & $34$ & $0.49$  \\\hline
\end{tabular}
\end{adjustbox}
\caption{\it Basic cuts and efficiencies (in percent) for the BP and the main background}\label{tab:cuts}
\end{center}
\end{table}

\begin{figure}[b]
 \includegraphics[width=\columnwidth]{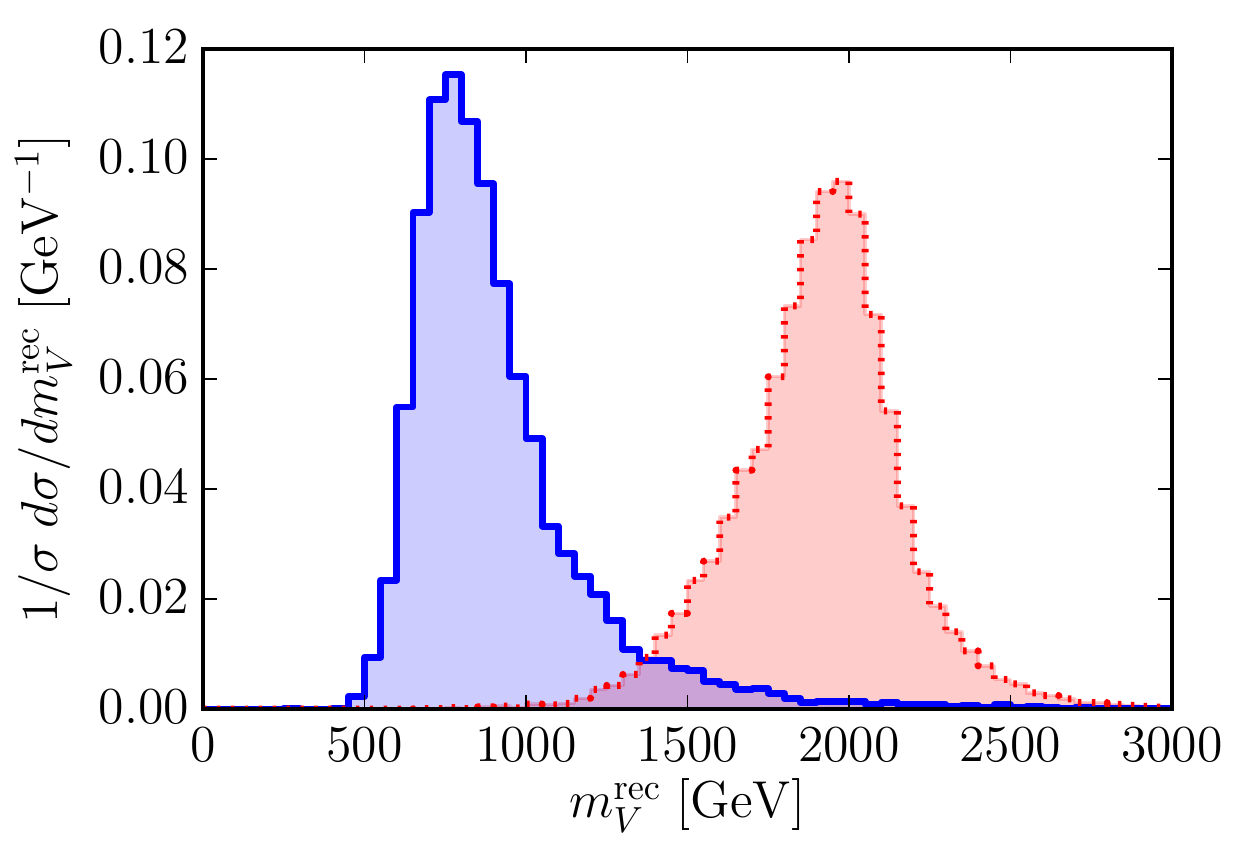}
 \caption{\it Normalized distribution of $m_V^\text{rec}$ in the BP (dashed red) and the background (solid blue).}\label{fig:massVp}
\end{figure}
\begin{figure}[t]
 \includegraphics[width=\columnwidth]{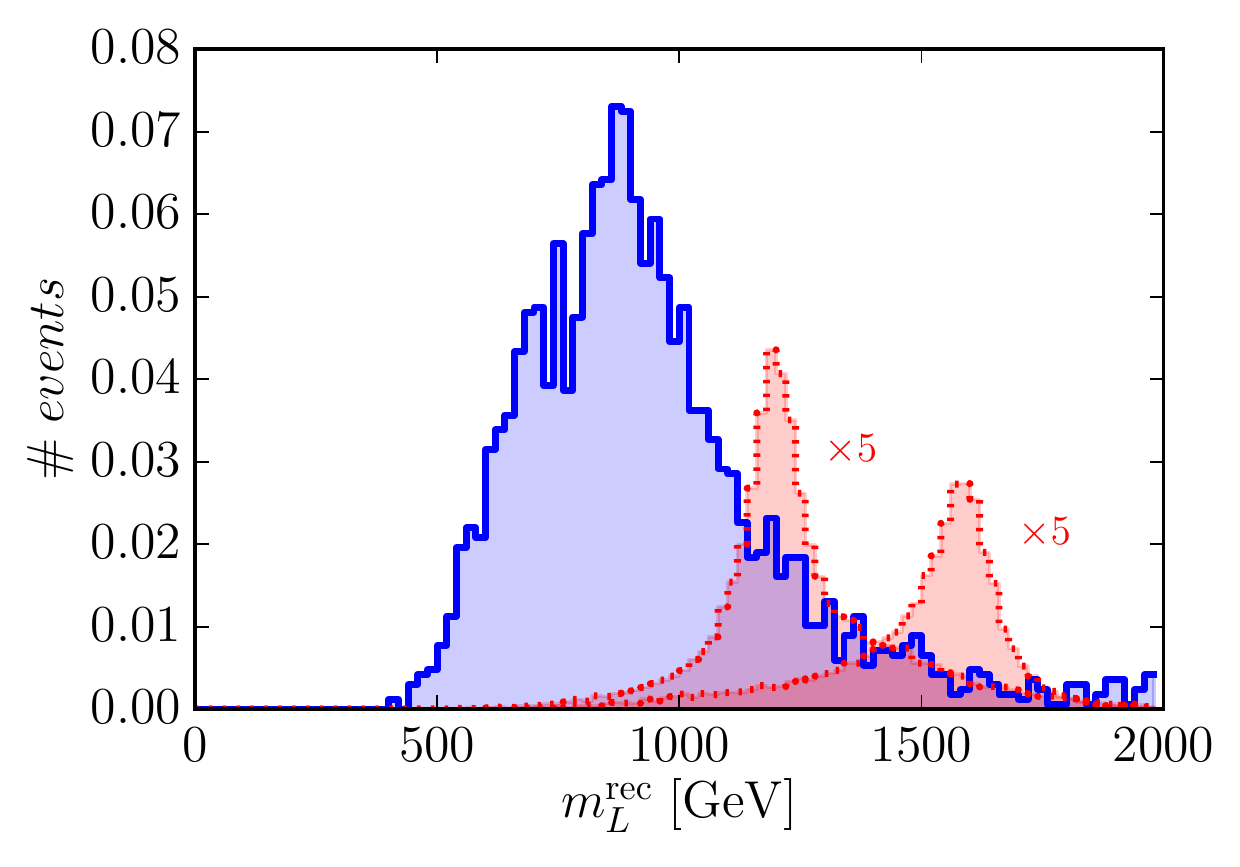}
 \caption{\it Event distribution of $m_L^\text{rec}$ for $\mathcal{L} = 1$ fb$^{-1}$ in the BP (dashed red) and in the background (solid blue). The BP with $m_L = 1.6$ TeV is also shown in red. Both signals are multiplied by a factor of $\times 5$ to increase the readability.}\label{fig:mL}
\end{figure}

\begin{figure*}[t]
\begin{center}
 \includegraphics[width=2.\columnwidth]{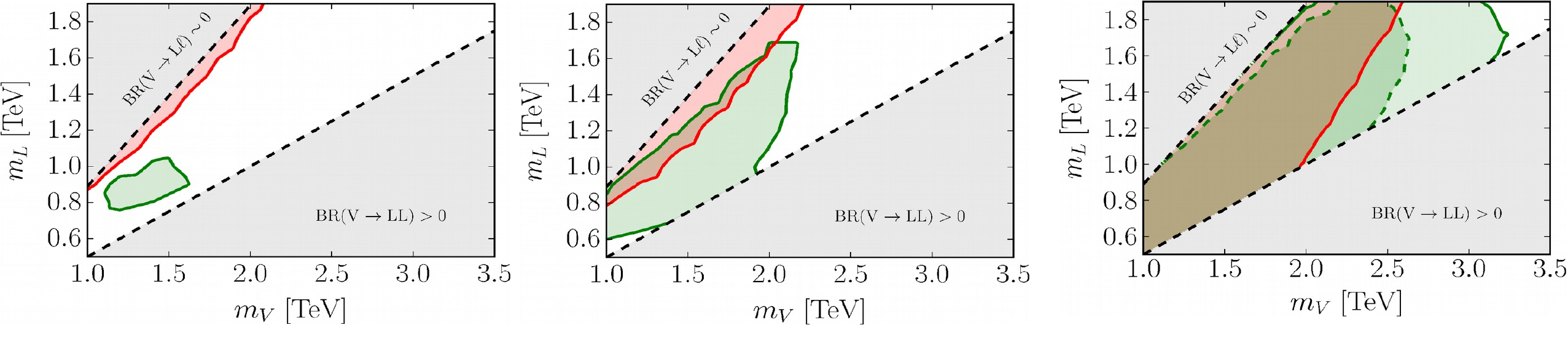}
 \end{center}
 \caption{\it Left) Parameter space region in the plane $m_V$--$m_L$ that can be excluded at the $95$ \% CL using di-muon searches (solid red) and using our dedicated analysis (solid green) for $\mathcal{L} = 70$ fb$^{-1}$. Center) Same as before but for $\mathcal{L} = 300$ fb$^{-1}$. Right) Same as before but for $\mathcal{L} = 3000$ fb$^{-1}$; the region enclosed by the dashed green line assumes half of the expected cross section into the target final state (due \textit{e.g.} to the presence of additional decay modes).}\label{fig:sig}
\end{figure*}

After these basic cuts, $m_V$ can be reconstructed just as the invariant mass $m_V^\text{rec} = m(\mu_1 + \mu_2 + j_1)$. The normalized distribution in the BP and in the background after the basic cuts is depicted in Fig.~\ref{fig:massVp}. Clearly,
a cut on $m_V^\text{rec}\gtrsim 1$ TeV separates further the signal from the SM.

For large $m_L$, the muon coming from $E$ is typically that with largest $p_T$, \textit{i.e.} $\mu_1$. $m_L$ can then be obtained just as the invariant mass $m_L^\text{rec} = m(\mu_1+j_1$). The corresponding distribution in the signal as well as in the background after the basic cuts and after enforcing $m_V^\text{rec} > 1$ TeV is depicted in Fig.~\ref{fig:mL}. 
We normalize to the expected number of events at $\mathcal{L} = 1$ fb$^{-1}$. Remarkably, the signal is already comparable to the background in number of events, while concentrating at much larger values of $m_L^\text{rec}$.

Given our ignorance on $m_V$ and $m_L$, we set further cuts depending on these parameters. In particular, we require $m_V^\text{rec} > 0.75\times m_V$, $|m_L^\text{rec}-m_L| < 100$ GeV. Denoting by $S$ and $B$ the number of signal and background events, respectively, after all cuts, we estimate the significance as $\mathcal{S} = \frac{S}{\sqrt{S+B}}$.
In Fig.~\ref{fig:sig} we show, in green, the regions that can be excluded at the $95\%$ CL ($\mathcal{S} = 2$) as a function of $m_L$ and $m_V$ for different values of the collected luminosity. These are to be compared with the reach of di-muon searches, depicted in red. In the grey regions, the composite spin-$1$ resonance does not decay sizeably into $L\ell$. The reason is, that in the upper triangle $L$ is heavier than (or very close in mass to) $V$. In the lower triangle, $V$ decays mostly into pairs of heavy leptons.

Remarkably though, regions not yet tested by the LHC can start to be probed. Moreover, contrary to di-muon searches, our new analysis will also shed light on the high mass region.
%

\section{Conclusions}~\label{ref:conclusions}
If the origin of the apparent breaking of Lepton Flavour Universality (LFU) in $B$ meson decays is due to a composite spin-$1$ resonance $V$, this has to couple to rather composite light leptons as well as quarks. We have shown that $V$ should not be searched in di-muon final states, as it has been done so far, but rather into di-top or in final states containing both a heavy and a light fermions. Focusing on the case of composite muons, we have discussed when $V$ decays mostly into a muon and a composite fermionic resonance, leading to a final state consisting of two muons and a boosted gauge or Higgs boson which express mainly as fat jets. Unravelling the physics responsible for breaking LFU in such a final state requires dedicated and tailored analyses to its kinematic features.

We have worked out one such analysis based on jet substructure techniques. Three main conclusions can be pointed out. \textit{(i)} Parameter space regions that were thought excluded by searches for di-muon resonances, are still allowed. They are not even ruled out by other beyond the Standard Model analyses, including multi-lepton searches for electroweakinos or heavy leptons.
\textit{(ii)} Some of the allowed regions can already be probed at the $95\%$ CL with our dedicated analysis. \textit{(iii)} With more luminosity, \textit{e.g.} $300$ ($3000$) fb$^{-1}$, heavier resonances can be tested, \textit{e.g} $m_V \sim 2$ ($3$) TeV.

Had we assumed $\lambda_{33}^\ell\sim\lambda_{22}^\ell$ in Eq.~\ref{eq:par}
 ~\footnote{In principle $\lambda_{11}^\ell$ could also be large~\cite{Carmona:2015ena,Carmona:2017fsn}, but the naive expectation in composite Higgs models is that composite particles couple stronger to heavier states.},
the cross section for our target final state would be reduced by an $\mathcal{O}(1)$ factor. The analogous topology with taus instead of muons would be also important. Not taking the latter into account, we estimate the reduced reach to this setup in the right panel of Fig.~\ref{fig:sig}; see dashed green line. Finally, composite electrons give signatures very similar to the ones studied here, with electrons instead of muons in the final state. Our analysis can therefore be trivially extended to this latter case, for which we expect roughly the same sensitivity.
On balance, we strongly encourage a re-analysis of current data. 

%

\acknowledgments
%
We acknowledge Shankha Banerjee for help with the analysis, and Kaustubh Agashe, Adri\'an Carmona, Luca Di Luzio and Mariano Quir\'os for useful discussions and comments on the manuscript. MC is supported by the Royal Society under the Newton International Fellowship programme. MC acknowledges the hospitality of the Weizmann Institute during the completion of this work.

\appendix

\bibliography{notes}{}

\end{document}